\documentclass{emulateapj} 
\usepackage{graphicx}
\usepackage{color}
\usepackage{dashrule}
\usepackage{multirow,bigstrut,bigdelim}
\usepackage{epsfig}

\begin{document}

\title{GRB 120422A: A Low-luminosity Gamma-ray Burst Driven by Central Engine}

\author{\sc Bin-Bin Zhang\altaffilmark{1,*}, Yi-Zhong Fan\altaffilmark{2,3,*}, Rong-Feng Shen \altaffilmark{4}, Dong Xu\altaffilmark{5,6}, Fu-Wen Zhang\altaffilmark{7}, Da-Ming Wei\altaffilmark{2,3}, David N. Burrows\altaffilmark{1}, Bing Zhang\altaffilmark{8,*} and Neil Gehrels\altaffilmark{9}}
\altaffiltext{1}{Department of Astronomy and Astrophysics, Pennsylvania State University, University Park, PA 16802, USA}
\altaffiltext{2}{Purple Mountain Observatory, Chinese Academy of Sciences, Nanjing 210008, China;}
\altaffiltext{3}{Key Laboratory of Dark Matter and Space Astronomy, Chinese Academy of Sciences, 210008, Nanjing, China;}
\altaffiltext{4}{Department of Astronomy \& Astrophysics, University of Toronto, M5S 3H4, Canada}
\altaffiltext{5}{Benoziyo Center for Astrophysics, Faculty of Physics, The Weizmann Institute of Science, Rehovot 76100, Israel}
\altaffiltext{6}{National Astronomical Observatories, Chinese Academy of Sciences, Beijing 100012, China}
\altaffiltext{7}{College of Science, Guilin University of Technology, Guilin, Guangxi 541004, China.}
\altaffiltext{8}{Department of Physics and Astronomy, University of Nevada Las Vegas, Las Vegas, NV 89154, USA}
\altaffiltext{9}{NASA Goddard Space Flight Center, Greenbelt, MD 20771, USA}
\altaffiltext{*}{Email: bbzhang@psu.edu (BBZ); yzfan@pmo.ac.cn (YZF); zhang@physics.unlv.edu (BZ).}
\begin{abstract}
GRB 120422A is a low-luminosity Gamma-ray burst (GRB) associated with a bright supernova, which distinguishes itself by its relatively short $T_{90}$ ($\sim 5$ s) and an energetic and steep-decaying X-ray tail. We analyze the Swift BAT and XRT data and discuss the physical implications. We show that the early steep decline in the X-ray light curve can be interpreted as the curvature tail of a late emission episode around 58-86 s, with a curved instantaneous spectrum at the end of the emission episode.
Together with the main activity in the first $\sim 20$ s and the weak emission from 40 s to 60 s, the prompt emission is variable, which points towards a central engine origin, in contrast to the shock breakout origin as invoked to interpret some other nearby low-luminosity supernova GRBs. The curvature effect model and interpreting the early shallow decay as the coasting external forward shock emission in a wind medium both give a constraint on the bulk Lorentz factor $\Gamma$ to be around several.
Comparing the properties of GRB 120422A and other supernova GRBs, we find that the main criterion to distinguish engine-driven GRBs from the shock breakout GRBs is the time-averaged $\gamma$-ray luminosity.  Engine-driven GRBs likely have a luminosity above $\sim 10^{48}~{\rm erg~s^{-1}}$.
\end{abstract}


\setlength{\parindent}{.25in}

\section{Introduction}
GRB 110422A triggered the Burst Alert Telescope (BAT; Barthelemy et al. 2005) on-board {\em Swift} at 07:12:03 UT on 2012 April 22 (Troja et al 2012). {\em Swift} slewed to the burst immediately. The two narrow field instruments, the X-ray Telescope (XRT; Burrows et al. 2005) and the Ultraviolet Optical telescope (UVOT; Roming et al. 2005) on-board {\em{Swift}} began to observe the field at $T_0+95.1 $ s and $T_0+104 $ s, respectively, where $T_0$ is the BAT trigger time. A bright X-ray afterglow was localized at ${\rm R.A. (J2000)}=09^h07^m38.46^s$, ${\rm Dec. (J2000)}=+14^{\circ}01'05\farcs 6$ with an uncertainty of 1\farcs9 (90\% confidence, Beardmore et al. 2012). A UVOT source was found within the XRT error circle (Kuin \& Troja 2012) and was confirmed by several ground follow-ups (e.g., Tanvir et al. 2012; Nardini et al 2012; Rumyantsev et al. 2012). A redshift $z=0.283$ was measured, and an associated supernova was soon discovered (Malesani et al 2012a, b; Melandri et al 2012; Wiersema et al. 2012; Sanchez-Ramirez et al 2012). This firmly places the burst in the massive-star core collapse category (Type II/Long, Zhang et al. 2009a). An unusual property of the burst is the large offset of the GRB position from the center of its host galaxy, which is often interpreted as evidence for a compact-star-merger origin (Type I/short) of the burst (had the associated SN not been discovered). This might be related to massive star formation/death in an interacting system (Tanvir et al 2012; Sanchez-Ramirez et al 2012).

In this paper, we focus on the early-time {\em Swift} data of this burst, aiming at understanding its physical origin. We present our data analysis of the BAT and XRT data in \S 2, and compare GRB 120422A with other SN-associated GRBs in \S 3. In \S 4, we then discuss the possible physical origins of prompt emission and early afterglow and constrain the bulk Lorentz factor.
The results are summarized in \S 5, along with a discussion on the physical implications of this event.

\section{Data analysis}

\begin{figure}
 \begin{center}
 \includegraphics[width=0.5\textwidth]{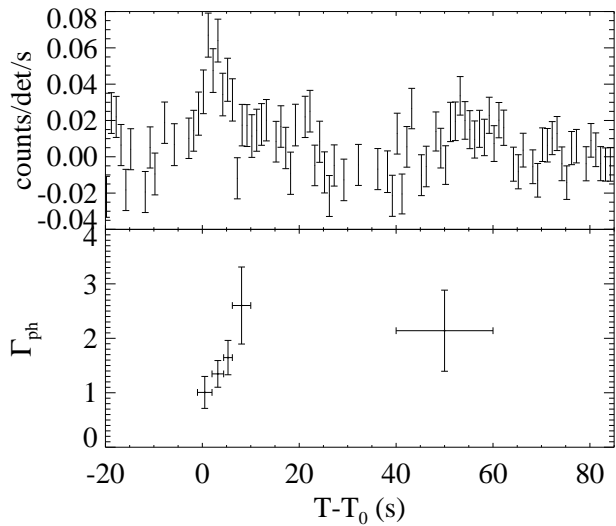}
 \end{center}
 \caption{The BAT count rate (upper panel) and photon index evolution (lower panel) of GRB 120422A. The spectral model is a simple power law  (``powerlaw" in {\it Xspec}).}
\label{fig:batlc}
\end{figure}

We processed the {\em Swift}/BAT data using standard {HEAsoft} tools (version 6.11). As shown in Fig. \ref{fig:batlc}, the main burst lasted from $T_0-3$ seconds to $T_0+20$ seconds with $T_{90}=5.4\pm 1.4$ seconds. We extracted the BAT spectra in five time slices. The lower panel in Fig. \ref{fig:batlc} shows the photon indices obtained by fitting the spectra with a simple power-law model. It is obvious that this burst has a strong hard-to-soft spectral evolution, which is similar to most other {\it Swift}/BAT GRBs. The photon indices range from $\sim 1.0$ to $\sim 2.6$. The time-integrated spectrum from $0-10$ s can be fitted with a simple power law with photon index $\Gamma_{\rm ph}=1.94\pm0.3$. Weak emission (at 3$\sigma$ level) was observed at $40-65$ s with a low-significance peak at $t\sim 45$ s and photon index $\sim 2.1\pm 0.7$. No significant pre-trigger emission was detected in the BAT band up to $T_0-200$ s.

The BAT band (15-150 keV) peak flux is 0.6$\pm$ 0.2 photons cm$^{-2}$ s $^{-1}$, and the total fluence is about $2.3\pm0.4 \times 10^{-7}\ {\rm erg} \ {\rm cm}^{-2}$. For a burst at a redshift $z=0.283$, this corresponds to a peak luminosity $L\sim 10^{49}$ erg s$^{-1}$ and total isotropic energy $\sim $ $4.5\times 10^{49}$ erg. The peak luminosity is well below the typical luminosity $\sim 10^{52}$ erg/s of bright GRBs, but is considerably higher than those of some nearby low-luminosity GRBs (e.g. $L\lesssim ~{\rm a~few\times}~10^{47}$ erg/s).

In a standard fashion, we processed the {\em Swift}/XRT data using our own IDL code which employs the standard HEAsoft analysis tools. For technical details please refer to Zhang et al. (2007a). Fig. \ref{fig:xrtlc} shows the XRT light curve and spectral evolution. The XRT light curve shows an unusually steep (decay slope $>$ 6) X-ray tail between $T_0+85$ s to $T_0+1000$ s, then followed by a shallow decay phase with decay slope $\sim 0.25$. A break is observed at $\sim 10^5$ s before the final normal decay phase (decay slope $\sim 1$). The X-ray spectrum can be fitted with an absorbed power-law. Strong spectral evolution was observed in the steep decay phase where the photon indices vary significantly from $\Gamma_{\rm ph} \sim 2.1$ to $\Gamma_{\rm ph} \sim 3.5$. The late time spectrum has no significant evolution with an average photon index $\Gamma_{\rm ph} \sim 2.1$. The total fluence in the XRT band (0.3-10 keV) is $1.53\pm0.26 \times 10^{-7} \ {\rm erg} \ {\rm cm}^{-2}$ (from $\sim 86.3$ s to $10^6$ s; corrected for XRT observation gaps).

\section{Comparison with other supernova GRBs}

{As shown in Table \ref{tab:1}, among the bursts associated with a well-monitored supernova, GRB 120422A distinguishes itself by the following facts: (1) it has the shortest $T_{90}$; (2) the initial luminosity of the X-ray radiation is high (e.g. greater than that of GRB 060218 and GRB 100316D by a factor of 100, see Fig.\ref{fig:snlc}) and the temporal decay slope is steep; (3) the X-ray afterglow plateau is also significantly brighter than GRB 060218 and GRB 100316D in the same time frame (i.e., $10^4$-$10^5$ s), even though the total prompt emission $\gamma$/X-ray energies of these three bursts are comparable. This suggests that a much higher energy is carried by the relativistic outflow in GRB 120422A. In fact, among the bursts with a well-monitored spectroscopic supernova detected so far (the ``Gold" sample in Table \ref{tab:1}), at one day after the burst, the X-ray afterglow of GRB 120422A is only dimmer than that of GRB 030329, a typical high luminosity GRB in the nearby universe. (4) There is a large offset ($\sim 8$ kpc; Tanvir et al 2012) between the burst location and the center of its host galaxy, which is rather unusual for massive star core-collapse GRBs (see e.g., Fruchter et al. 2006; Zhang et al 2009a).
Within the Gold sample of supernova GRBs, the isotropic gamma-ray energy $E_{\gamma,iso}$ of GRB 031203, GRB 060218, GRB 100316D, GRB 120422A are rather similar. Interestingly they seem to belong to two sub-classes. As already noticed earlier (e.g., Fan et al 2011; Starling et al. 2011), XRF 060218 and XRF 100316D are cousins, since both their spectral and temporal behaviors are rather similar (see also Fig.\ref{fig:snlc}), except that the former was associated with a less energetic SN 2006aj. On the other hand, GRB 120422A and GRB 031203 share quite a few similarities. For example, they are both relatively short;
their peak luminosities during the prompt emission phase are almost identical; their 15-150 keV spectra are both soft with spectral indices $\alpha \sim 0.6-0.9$ ($\alpha$ is defined as $f_\nu \propto \nu^{-\alpha}$); and their late-time ($t>1$ day) X-ray afterglow luminosities are comparable with each other, but are significantly brighter than GRB 060218 and XRF 100316D.


\begin{deluxetable}{llllll}
 \tablecaption{The observational properties of GRB 120422A and other supernova GRBs}
  
 \tablecolumns{1}
 \tabletypesize{\footnotesize}
 \renewcommand{\tabcolsep}{0.05cm}
\startdata

   \hline\noalign{\smallskip}
GRB & $z$ & $T_{90}$ & $E_{\rm peak}$ &$E_{\rm \gamma,iso}$       &  Ref.\tablenotemark{b}\\

   &      &   (s)     &   (keV)       & ($10^{51}$ erg)           &     \\
  \hline\noalign{\smallskip}

\multicolumn{6}{c}{Gold\footnote{The gold sample includes Type II GRBs that have spectroscopically 
identified supernova association, and also well-monitored supernova emission. 
The Silver sample includes GRBs that have a clear supernova bump along with some 
spectroscopic evidence. The similar categorization was also adopted by Hjorth 
\& Bloom (2011).}} \\

\multicolumn{6}{c}{\hdashrule{90mm}{0.5pt}{1pt}} \\
980425  &0.0085 &34.9$\pm$3.8 &122$\pm$17             &$9\times10^{-4}$    & 1,2\\
030329  &0.1685 &22.9         &70$\pm$2               &13                & 1\\
031203  &0.1055 &37.0$\pm$1.3 &$>190$                 &0.17                  &1,3 \\
060218  &0.0334 &2100$\pm$100 &4.7$\pm$1.2            &0.04              & 1,2\\
100316D &0.0591 &$>$1300      &18$^{+3}_{-2}$         &0.06                  & 1,4 \\
120422A &0.283  &5.35$\pm$1.4 & $\sim 53$\tablenotemark{a} &0.045         & 5,6 \\
 \hline\noalign{\smallskip}
\multicolumn{6}{c}{Silver$^a$} \\

\multicolumn{6}{c}{\hdashrule{90mm}{0.5pt}{1pt}} \\
011121  & 0.362 & $\sim 28$  &     & 27 &      7\\
020903  & 0.251 & $\sim 20$  & $\sim 2 $     & 0.011   & 8,9 \\
021211  & 1.006 &  $\sim 4$ &   $46.8^{+5.8}_{-5.1}$   & 6.6        &  10\\
050525A  & 0.606 &  $8.8\pm 0.5$  &  $84.1\pm1.7$  & 23        &  11\\
081007 & 0.5295  & 8  & $61\pm 15$   & 1.5    &     12\\
101219B & 0.55 &  51  & $70\pm 8$    & 4.2 &        13 
\enddata
\tablenotetext{b}{References: [1] Hjorth \& Bloom (2011); [2] Zhang et al. (2008); [3] Sazonov et al. (2004); [4] Sakamoto et al. (2010); [5] Barthelmy et al. (2012); [6] Schulze et al. (2012); [7] Garnavich et al. (2003); [8] Sakamoto et al. (2004); [9]  Soderberg et al. (2004); [10] Crew et al. (2003); [11] Blustin et al. (2006); [12] Jin et al. (2012); [13] Sparre et al. (2011).}
\label{tab:1}
\end{deluxetable}

\begin{figure}
 \begin{center}
 \plotone{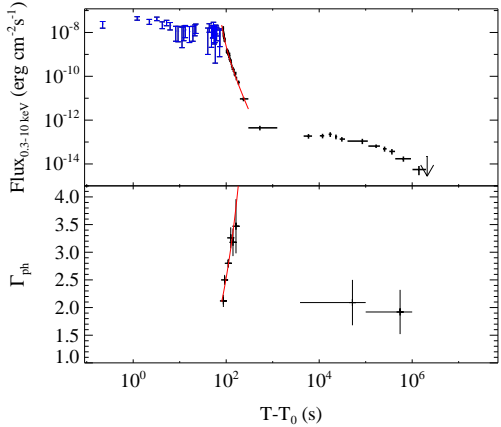}
 \end{center}
\caption{{\it Upper panel:} The {\textit Swift}/XRT light curve of GRB 120422A (black) and the BAT light curve extrapolated to the XRT band (blue). The solid red lines show the curvature effect model (Zhang et al. 2009b) fitted to the observed flux. {\it Lower panel}: photon index evolution. The solid red line shows the curvature effect model (Zhang et al. 2009b) fitting to the observed photon index. See \S 4 for details.}
\label{fig:xrtlc}
\end{figure}

\begin{figure}
 \begin{center}
 \plotone{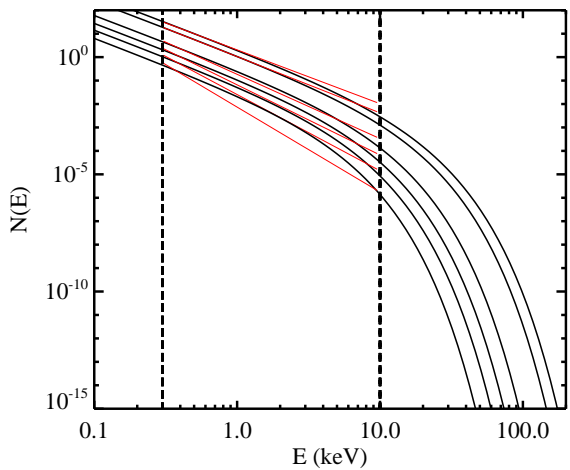}
 \end{center}
\caption{Time-dependent theoretical spectra based on the curvature effect of a non-power-law spectrum. From top to bottom, each spectrum corresponds to a time slice of the steep decay phase, which is the same as that in the lower panel of Fig. \ref{fig:xrtlc}. The XRT band (0.3-10 keV) is bracketed by two vertical lines. The red solid lines show the effective power-law model in the narrow XRT band.}
\label{fig:spectrum}
\end{figure}

\section{X-ray afterglow modeling and engine-driven GRB}\label{sec:X-ray-tail}

\subsection{The steep decay phase}

The steep decay phase is commonly observed in Swift GRBs (e.g. Tagliagerri et al. 2005; Barthelmy et al. 2005). The standard interpretation of this phase is the ``curvature'' tail of the prompt emission (Fenimore et al. 1996; Kumar \& Panaitescu 2000; Zhang et al. 2006; Liang et al. 2006; Zhang et al. 2009b), which arises from delayed photon emission from high latitudes with respect to the line of sight upon the abrupt cessation of the prompt emission. Other interpretations include rapid expansion of a thermal plasma associated with shock breakout  or a hot cocoon surrounding a jet after exiting the progenitor star (e.g. Fan et al. 2006; Pe'er et al. 2006).

In the shock breakout picture (Fan et al. 2006), a quick decline in X-rays is possible for a quasi-thermal spectrum $F_{\nu_{\rm obs}}\propto R^{2}e^{-h\nu_{\rm obs}/kT_{\rm obs}}$ for $h\nu_{\rm obs}\gg kT_{\rm obs}$. The temperature drops with time as $T_{\rm obs}\propto R^{-a/3}$, where $a=2$ if the width of the hot material is fixed, or $a=3$ if the width of the hot material is proportional to the radius $R$. Taking $a=3$ as an example, i.e., $T_{\rm obs} \propto R^{-1}\propto t^{-1}$, the XRT-band luminosity can be expressed as
\begin{equation}
L_{\rm XRT}\propto \int^{10~\rm keV}_{\rm 0.3~keV}F_{\nu_{\rm obs}}d\nu_{\rm obs}\propto F(t)e^{-At},
\end{equation}
where $A>0$ is a constant and $F(t)>0$ is a function of $t$. In principle, this model can give rise to a progressively steepening steep-decay phase with rapid spectral evolution (for $kT < 0.3$ keV). This model does not fit the data. Also the emergence of a shallow decay component is not expected within such a scenario.

We then investigate the curvature effect model for a non-power-law spectrum (Zhang et al. 2009b).
We consider a time-dependent cut-off power-law photon spectrum taking the form
\begin{equation}
N(E,t)=N_0(t)\left(\frac{E}{1~{\rm keV}}\right)^{-\Gamma_{\rm ph}}e^{-\frac{E}{E_c(t)}},
\end{equation}
where $\Gamma_{\rm ph}$ is the power-law photon spectral index, $E_c(t)=E_{c,p}[(t-t_0)/(t_p-t_0)]^{-1}$ is the time-dependent characteristic cut-off photon energy, $N_0(t)=N_{0,p}[(t-t_0)/(t_p-t_0)]^{-(1+\Gamma_{\rm ph})}$ is a time-dependent photon flux, and $t_0$ refers to the time origin of the last/main pulse in the prompt emission. Denoting $t_p$ as the peak time of the last pulse, one can derive time-dependent decay index and the effective spectral index using the formalism derived in Zhang et al. (2009b). For GRB 120422A, $t_p$ cannot be inferred from the XRT light curve, since the X-ray already entered the steep decay phase when the XRT slewed to the source. To constrain $t_p$, we extrapolate the BAT flux to the XRT band assuming a simple power law model extending all the way to the XRT band.
It is found that the extrapolated BAT flux to the XRT band and the observed XRT light curve intersect around the time when the XRT observation started (Fig.\ref{fig:xrtlc}). We thus take $t_p \sim 86.3$ s (the beginning of XRT observation) in our modeling.

We successfully fit both the observed light curve and the photon index curve with our model, and get the following best-fit parameters: $N_{0,p}=2.36\pm0.09$, $E_{c,p} = 7.62_{-0.83}^{+0.97} $ keV, $\Gamma_{\rm ph} = 2.30 \pm 0.07 $,  $t_0 = 57.5\pm 0.65$ s, with $\chi^2/dof = 71.2/52$.
(Fig.\ref{fig:xrtlc}). Figure \ref{fig:spectrum} gives the modeled spectra as a function of time.
This suggests that there was likely a central-engine-powered emission in the time interval $58~{\rm s} - 86~{\rm s}$. Together with the main activity in the first $\sim 20$ s and the weak/soft radiation from 40 s to 60 s, the variability of the prompt emission of GRB 120422A is well established. This strongly favors a central engine origin of the observed prompt emission.

For high latitude emission, a rough constraint on the emission radius, and hence, bulk Lorentz factor $\Gamma$ of the outflow (within the framework of the internal shock model) may be imposed (e.g. Zhang et al. 2006; Jin et al. 2010).
The length of the tail emission $t_{\rm tail} = t - t_p$ can be expressed as $t_{\rm tail} \leq 2 \Gamma^2 \Delta t (1-\cos\theta_j)$, where $\Delta t \sim t_p - t_0$ is the variability time scale, and $\theta_j$ is the jet opening angle. Plugging in the numbers, i.e. $t_{\rm tail} \sim 250-86 = 164$ s, $\Delta t = 29$ s, one can derive a constraint
\begin{equation}
 \Gamma \geq \frac{1.68}{\sqrt{1-\cos\theta_j}}.
\end{equation}
For $\theta_j= 10^\circ, 20^\circ, 30^\circ$, the corresponding constraints are
$\Gamma \geq 13.6, 6.8, 4.6$, respectively.

\subsection{The plateau phase}

Following the steep decay phase is an X-ray plateau, lasting until $\sim$ 1 day after trigger. This component is commonly observed in high-luminosity GRBs, and there is no consensus regarding its interpretation. We discuss the following two possible interpretations:

{\bf \emph{Scenario I}:} The X-ray plateau is due to the forward shock emission of a mildly relativistic outflow during the ``coasting phase'' before significant deceleration starts (e.g. Shen \& Matzner 2012). For a wind medium with
density profile $n=3\times 10^{35}~{\rm cm^{-2}}~A_* r^{-2}$, one can show that the decay rate is very shallow in this phase, i.e. $F_\nu \propto t^{-(p-2)/2}\propto t^{-(\beta-1)}$, if the X-ray band frequency satisfies $\nu > \max(\nu_m,\nu_c)$.
The post-deceleration decay behavior in the same spectrum regime is $F_\nu \propto
t^{-(3p-2)/4} \propto t^{-(3\beta-1)/2}$. Both behaviors are in agreement with the data.

This interpretation leads to the following constraints: (1) the outflow deceleration time $t_{\rm dec}= t_b$, where $t_b=10^5$ s is the shallow-to-normal break time; (2) the external forward shock flux density at $t_b$ is as measured, $F_{\nu_x}(t_b)= 1.25\times10^{-2}$ $\mu$Jy; (3) $\nu_m(t_1) \leqslant \nu_x$; and (4) $\nu_c(t_2) \leqslant \nu_x$, where $t_1= 10^3$ s and $t_2= 10^6$ s are the observed starting time of the shallow decay and the lower limit of the end time of normal decay, respectively. 
The last two constraints are set in order to satisfy the spectral regime requirement $\nu>\max(\nu_m,\nu_c)$ for both the plateau and the normal decay phase, and are utilizing the model prediction that $\nu_m(t)$ decreases and $\nu_c(t)$ increases both with $t$ monotonically. We follow the formulae in Shen \& Matzner (2012, Eqs. 14-17 therein) which are based on the standard external shock synchrotron emission calculation (e.g., Sari et al. 1998) and include the numerical correction factor due to internal structure of shock and the equal-arrival-time surface (Granot et al. 1999). We adopt $\nu_x= 1$ keV and use $\beta= 2.1$ as observed.

Constraint (2) gives the wind medium density normalization 
\begin{equation} \label{eq:A_*}
A_*=0.4 \epsilon_e^{-1.14}\epsilon_B^{-0.05}\Gamma^{-4}
\end{equation}
where $\Gamma$ is the initial Lorentz factor of the outflow, and $\epsilon_e$ and $\epsilon_B$ are the shock
s electron and magnetic equipartition parameters, respectively. Combining constraints (1) and (2) gives the isotropic equivalent kinetic energy of the outflow
\begin{equation}    \label{eq:Ekiso}
E_{\rm k,iso}= 1.2 \times10^{51}~ \left(\frac{\epsilon_e}{0.01}\right)^{-1.14} \left(\frac{\epsilon_B}{0.01}\right)^{-0.05}~ \mbox{erg},
\end{equation}

Constraint (3) is trivial and easily satisfied. Utilizing Eq. (\ref{eq:A_*}), constraint (4) gives 
\begin{equation}
\Gamma \leqslant 6.5~ \left(\frac{\epsilon_e}{0.01}\right)^{-0.21} \left(\frac{\epsilon_B}{0.01}\right)^{0.18},
\end{equation}
This constraint is consistent with the curvature effect constraints if $\theta_j > 20^\circ$. So the entire afterglow data are consistent with a wide jet with a moderately high Lorentz factor $\Gamma \sim 6$.

{\bf \emph{Scenario II}:} If the jet is narrower, say $\theta_j < 20 ^\circ$, the X-ray plateau cannot be interpreted as the pre-deceleration forward shock in a wind medium. The deceleration time has to be much earlier, and the extended plateau can be interpreted as forward shock emission with significant energy injection\footnote{An alternative solution is to explain plateau phase as late prompt emission (see, e.g., Ghisellini et al.  2007).} (e.g. Zhang et al. 2006 and the references therein). There are two possible cases. Case (A): One can argue that the central engine is a millisecond magnetar with a dipole radiation luminosity $\sim 10^{47}~{\rm erg~s^{-1}}$ and a spin down timescale $\tau_0 \sim 10^{5}$ s. This gives a constraint on the surface magnetic field $B_p = (0.5-1)\times 10^{14}$ Gauss and the initial spin period $P_0 \sim 1$ ms. One potential challenge of this scenario is that the efficiency of the forward shock radiation in XRT band has to be extremely small (say, very low $\epsilon_e$). Otherwise, the resulting X-ray emission would be much brighter than what is observed.
Case (B): One may argue that the outflow has a Lorentz factor distribution and the distribution satisfies $E(>\Gamma)\propto \Gamma^{-5}$.

In both scenarios, the X-ray flux at $t \sim 10^5$ s constrains the total kinetic energy of the outflow at that time, which is given by Equation (\ref{eq:Ekiso}). However, the total kinetic energy of the initial outflow that produces the prompt burst, $E_{\rm k,p,iso}$, is different for the two scenarios. In scenario I, $E_{\rm k,p,iso}= E_{\rm k,iso}$, while in scenario II, $E_{\rm k,p,iso} \ll E_{\rm k,iso}$.

\section{Conclusions and Discussion}

We have analyzed the BAT and XRT data of the nearby, low-luminosity, supernova-associated GRB 120422A. Even though $T_{90}$ of the burst is short, BAT emission shows extended fluctuation signals, suggesting a possible extended central engine activity. This is confirmed by the XRT data, which showed a rapid decline followed by an extended plateau similar to most other high-luminosity GRBs. The rapid decline tail can be modeled by the curvature effect model of Zhang et al. (2009b). The derived beginning time of the last emission episode is about 58 s, with the last peak near 86 s. Various arguments (see below for more discussion) suggest that this low-luminosity GRB is central-engine-driven, rather than powered by shock breakout. The Lorentz factor of the ejecta is constrained to be at least moderately relativistic. 

As discussed above, an engine-driven origin is supported by the following facts: (1) The $\gamma$-ray light curve is variable; (2) the rapidly decaying prompt tail emission is inconsistent with a cooling thermal emission component from shock breakout, but is consistent with the curvature tail of a successful jet; and (3) a long lasting X-ray shallow decay followed by the steep decay is consistent with external shock emission of a successful jet.

Some nearby low-luminosity GRBs may have signature of shock breakout (e.g. GRB 060218, Campana et al. 2006; Waxman et al. 2007, but see Ghisellini et al. 2006, 2007; Li 2007; Bjornsson 2008; Chevalier \& Fransson 2008; Page et al. 2011). The event rate of nearby low-luminosity GRBs is much higher than the simple extrapolation of the high-luminosity GRB event rate, making a distinct population (e.g. Soderberg et al. 2006; Liang et al. 2007; Virgili et al. 2009; Coward 2005). Some authors have suspected that low-luminosity GRBs may be unsuccessful jets, and the radiation signal is mostly powered by shock breakout. The relativistic shock break out model predicts a ``fundamental plane'' correlation  $T_{90}\sim 20~{\rm s}~(1+z)^{-1.68}\left(\frac{E_{\rm \gamma,iso}}{10^{46}~{\rm erg}}
\right)^{1/2}\left(\frac{E_{\rm p}}{50~{\rm keV}}\right)^{-2.68}$ (Nakar \& Sari 2012). For the parameters of this burst, $E_{\rm \gamma,iso} \sim  4\times 10^{49}$ erg and $E_{\rm p}\sim 53$ keV, the predicted shock breakout duration is $\sim 1100$ s, much longer than $T_{90} \sim 5$ s, or the extended duration $\sim 86$ derived from the curvature effect fitting. This is a strong evidence against the shock breakout interpretation of this burst.

\begin{figure}
 \begin{center}
 \includegraphics[width=0.45\textwidth]{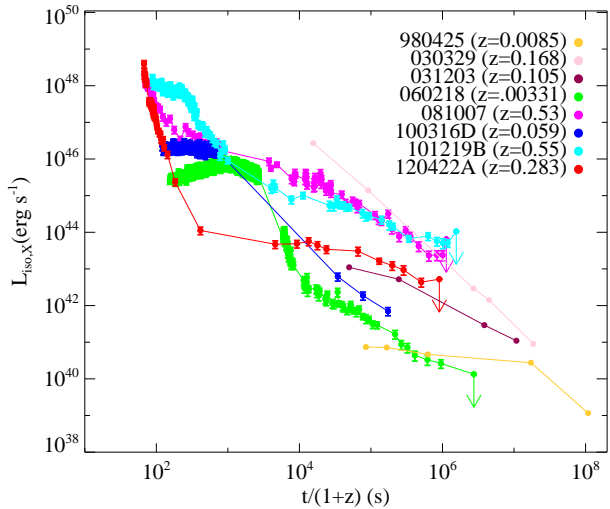}
 \end{center}
\caption{A comparison of the observed X-ray luminosity light curves of GRB 120422A and other supernova GRBs. The data of GRB 980425, GRB 030329, GRB 031203, GRB 060218, XRF 100316D are the same as that of Fig.2 of Fan et al. (2011). The data of GRB 081007, GRB 101219B and GRB 120422A are analyzed in this work.}\label{fig:snlc}
\end{figure}

\begin{figure}
 \begin{center}
 \includegraphics[width=0.53\textwidth]{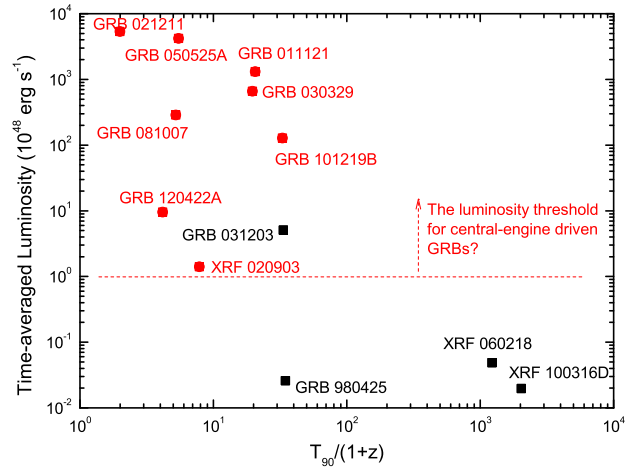}
 \end{center}
\caption{Supernova-associating GRBs in the time-averaged luminosity - $T_{90}/(1+z)$ plane. Red symbols denote engine-driven GRBs, while black ones denote the possible shock-breakout GRBs suggested in some literature. The red dashed line ($10^{48}~{\rm erg~s^{-1}}$) gives a rough threshold above which successful jet is possible.}
\label{fig:engine}
\end{figure}

In the collapsar model for GRBs, in order to make a successful jet, the central engine has to be active for a duration longer than the time for the jet to penetrate the star before breaking out. Otherwise the jet would be choked inside the star or quickly spread out upon the breakout. Considering the collimation of the jet by a surrounding cocoon, Bromberg et al. (2011) estimate the breakout time as
\begin{eqnarray}
t_{\rm B} & \simeq & 15~ \epsilon_{\gamma}^{1/3} \left(\frac{L_{\gamma,iso}}{10^{50} erg/s}\right)^{-1/3} \left(\frac{\theta_0}{10^{\circ}}\right)^{2/3} \\ \nonumber
          &        & ~~\times \left(\frac{R_*}{10^{11} cm}\right)^{2/3} \left(\frac{M_*}{15 M_{\odot}}\right)^{1/3}~ s,
\end{eqnarray}
where $\epsilon_{\gamma}$ is the burst radiation efficiency, and $\theta_0$ is the initial opening angle of the jet when it is injected from the central engine. Statistically, one would expect that the observed burst duration to be comparable or longer than this duration. For GRB 120422A, even if $T_{90} \sim 5$ is shorter than this jet penetration time, the real duration of the successful jet is actually near 86 s, as is constrained by the curvature effect modeling. The jet breakout condition is therefore satisfied.

What is the separation line between the engine-driven and shock-break GRBs? In Fig.\ref{fig:engine} all the supernova GRBs are plotted in the plane of time-averaged luminosity and $T_{90}$. It is shown that above $\sim 10^{48}~{\rm erg~s^{-1}}$, an engine-driven GRB is possible. Shock breakout luminosity cannot be much higher than this value. GRB 120422A belongs to the low end of engine-driven GRBs.

How could a successful GRB jet have such a low luminosity? The first
possibility may be related to its relatively low Lorentz factor (for the
scenario I of plateau interpretation). If this burst satisfies the empirical $\Gamma-E_{\rm \gamma,iso}$ and $\Gamma-L_{\gamma,iso}$ relations (Liang et al. 2010; L\"u et al. 2012; Fan et al. 2012), one would expect a moderately low $\Gamma$. This is generally consistent with the model constraints of $\Gamma$. 
Low-$\Gamma$ outflows tend to have low emissivities.
This can be due to an intrinsically low wind luminosity, or a smaller radiation
efficiency for an otherwise normal wind luminosity. This second possibility can
be related to internal shock model when the relative Lorentz factor between the
colliding shells is small (e.g. Barraud et al. 2005). Alternatively, the low
luminosity can be related to the viewing angle effect. A low-luminosity GRB can be obtained by an observer viewing at a large angle
from the jet axis of a structured jet (e.g. Zhang et al. 2004a). This may be
relevant for a hot cocoon surrounding a successful jet (e.g. Zhang et al.
2004b), which is consistent with low-$\Gamma$, large $\theta_j$ scenario
discussed in this paper. The scenario can be tested with the late-time radio
observations, which would give a more robust measure of the total energetics of
the event.

\acknowledgments
We thank Derek B. Fox, P\'eter M\'esz\'aros, Dirk Grupe and P\'eter Veres for helpful discussions. This work is supported in part by NASA SAO SV4-74018 (BBZ), the National Natural Science Foundation of China (grants 10973041, 10921063, 11073057 and 11163003) and National Basic Research Program of China under grant 2009CB824800, the 100 Talents Program of Chinese Academy of Sciences (YZF), NASA NNX10AD08G and NSF AST-0908362 (BZ).
\ 
\\


\begin{thebibliography}{}

\bibitem[Barraud et al.(2005)]{2005A&A...440..809B} Barraud, C., Daigne, F., Mochkovitch, R., \& Atteia, J.~L.\ 2005, \aap, 440, 809
\bibitem[Barthelmy et al. (2005)]{Barthelmy05} Barthelmy, S. D. et al. 2005, Space Science Reviews, 120, 143
\bibitem[Barthelmy et al.(2012)]{2012GCN..13246...1B} Barthelmy, S.~D., Baumgartner, W.~H., Cummings, J.~R., et al.\ 2012, GCN. Circ. 13246
\bibitem[Beardmore et al. (2012)]{Beardmore2012} Beardmore, A. P., Evans, P. A.,  Goad, M. R., \&  Osborne, J. P. 2012, GCN. Circ. 13247
\bibitem[Bj{\"o}rnsson(2008)]{2008ApJ...672..443B} Bj{\"o}rnsson, C.-I.\ 2008, \apj, 672, 443 
\bibitem[Blustin et al.(2006)]{2006ApJ...637..901B} Blustin, A.~J., Band, D., Barthelmy, S., et al.\ 2006, \apj, 637, 901
\bibitem[Bromberg et al (2011)]{bromber11} Bromberg, O., Nakar, E., Piran, T., Sari, R., 2011, ApJ, 740, 100
\bibitem[Burrows et al. (2005)]{Burrows05} Burrows, D. N., et al., 2005, Space Science Reviews, 120, 165
\bibitem[Campana et al. (2006)]{Campana06} Campana, S. et al., 2006, Nature, 442, 1008
\bibitem[Chevalier \& Fransson(2008)]{2008ApJ...683L.135C} Chevalier, R.~A., \& Fransson, C.\ 2008, \apjl, 683, L135 
\bibitem[Coward et al. (2005)]{Coward05} Coward, D. M. 2005, MNRAS, 360, L77
\bibitem[Crew et al.(2003)]{2003ApJ...599..387C} Crew, G.~B., Lamb, D.~Q., Ricker, G.~R., et al.\ 2003, \apj, 599, 387
\bibitem[Fan et al. (2006)]{Fan06} Fan, Y. Z., Piran, T., \& Xu, D. 2006, JCAP, 09, 013
\bibitem[Fan et al. (2011)]{Fan2011} Fan, Y. Z., Zhang, B. B., Xu, D., Liang, E. W., \& Zhang, B. 2011, ApJ, 726, 32
\bibitem[Fan et al.(2012)]{2012arXiv1204.4881F} Fan, Y.-Z., Wei, D.-M., Zhang, F.-W., \& Zhang, B.-B.\ 2012, arXiv:1204.4881 
\bibitem[Fenimore et al. (1996)]{Fenimore1996} Fenimore, E. E., Madras, C. D., \& Nayakshin, S. 1996, ApJ, 473, 998
\bibitem[Fruchter et al.(2006)]{2006Natur.441..463F} Fruchter, A.~S., Levan, A.~J., Strolger, L., et al.\ 2006, \nat, 441, 463
\bibitem[Garnavich et al.(2003)]{2003ApJ...582..924G} Garnavich, P.~M., Stanek, K.~Z., Wyrzykowski, L., et al.\ 2003, \apj, 582, 924
\bibitem[Ghisellini et al.(2007)]{2007MNRAS.375L..36G} Ghisellini, G., Ghirlanda, G., \& Tavecchio, F.\ 2007, \mnras, 375, L36 
\bibitem[Ghisellini et al.(2007)]{2007ApJ...658L..75G} Ghisellini, G., Ghirlanda, G., Nava, L., \& Firmani, C.\ 2007, \apjl, 658, L75 
\bibitem[Granot et al.(1999)]{1999ApJ...513..679G} Granot, J., Piran, T., \& Sari, R.\ 1999, \apj, 513, 679 
\bibitem[Hjorth \& Bloom (2011)]{Hjorth2011} Hjorth, J., \& Bloom, J. S. 2011, arXiv:1104.2274
\bibitem[Jin et al. (2010)]{Jin2010} Jin, Z. P., Fan, Y. Z., \& Wei, D. M. 2010, ApJ, 724, 861
\bibitem[Jin et al. (2012)]{Jin2012} Jin, Z. P. et al. 2012, ApJ, to be submitted.
\bibitem[Kumar \& Panaitescu (2000)]{Kumar2000} Kumar, P., \& Panaitescu, A. 2000, ApJ, 541, L51
\bibitem[Kuin \& Troja (2012)]{Troja2012} Kuin, N. P. M. \& Troja, E. 2012, GCN Circ. 13248
\bibitem[Li(2007)]{2007MNRAS.375..240L} Li, L.-X.\ 2007, \mnras, 375, 240 
\bibitem[Liang et al.(2007)]{Liang07} Liang, E., Zhang, B., Virgili, F., Dai, Z. G. 2007, ApJ, 662, 1111
\bibitem[Liang et al.(2010)]{Liang10} Liang, E.-W. et al. 2010, ApJ, 725, 2209
\bibitem[Lu et al.(2012)]{Lu12} L\"u, J. et al. 2012, ApJ, 751, 49
\bibitem[Malesani et al. (2012a)]{Malesani2012a} Malesani, D. et al. 2012a, GCN Circ. 13275
\bibitem[Malesani et al. (2012b)]{Malesani2012b} Malesani, D. et al. 2012b, GCN Circ. 13277
\bibitem[Mazzali et al. (2006)]{Mazzali06} Mazzali, P. A. et al., 2006, ApJ, 645, 1323
\bibitem[Melandri et al.(2012)]{2012arXiv1206.5532M} Melandri, A., Pian, E., Ferrero, P., et al.\ 2012, arXiv:1206.5532 
\bibitem[Nakar \& Sari (2011)]{Nakar2011} Nakar, E., \& Sari, R. 2012, ApJ, 747, 88
\bibitem[Nardini et al. (2012)]{Nardini2012} Nardini, M., Schmidl, S.,Greiner, J., \& Kann, D. A. 2012, GCN Circ. 13256
\bibitem[Page et al.(2011)]{2011MNRAS.416.2078P} Page, K.~L., Starling, R.~L.~C., Fitzpatrick, G., et al.\ 2011, \mnras, 416, 2078 
\bibitem[Pe'er et al. (2006)]{Peer2006}  Pe'er, A., M\'esz\'aros, P., Rees, M. J. 2006, ApJ, 652, 482
\bibitem[Roming et al. (2005)]{Roming2012} Roming, P. W., et al. 2005, Space Science Reviews, 120, 95
\bibitem[Rumyantsev et al. (2012)]{Rumyantsev2012} Rumyantsev, V., Antonyuk, K., \& Pozanenko, A. 2012, GCN Circ. 13273
\bibitem[Sakamoto et al.(2004)]{2004ApJ...602..875S} Sakamoto, T., Lamb, D.~Q., Graziani, C., et al.\ 2004, \apj, 602, 875
\bibitem[Sakamoto et al.(2010)]{2010GCN..10511...1S} Sakamoto, T., Barthelmy, S.~D., Baumgartner, W.~H., et al.\ 2010, GCN. Circ. 10511
\bibitem[Sanchez-Ramirez et al. (2012)]{Sanchez-Ramirez2012} Sanchez-Ramirez, R. et al., 2012, GCN Circ. 13281
\bibitem[Sari et al.(1998)]{1998ApJ...497L..17S} Sari, R., Piran, T., \& Narayan, R.\ 1998, \apjl, 497, L17 
\bibitem[Schulze et al.(2012)]{2012GCN..13257...1S} Schulze, S., Levan, A.~J., Malesani, D., et al.\ 2012, GCN. Circ. 13257
\bibitem[Sazonov et al. (2004)]{Sazonov2004} Sazonov, S. Y., Lutovinov, A. A., \&  Sunyaev, R. A. 2004, Nature, 430, 646
\bibitem[Shen \& Matzner (2012)]{SC12} Shen, R. F., Matzner, C. D., 2012, ApJ, 744, 36
\bibitem[Soderberg et al.(2004)]{2004ApJ...606..994S} Soderberg, A.~M., Kulkarni, S.~R., Berger, E., et al.\ 2004, \apj, 606, 994
\bibitem[Soderberg et al.(2006)]{2006Natur.442.1014S} Soderberg, A.~M., Kulkarni, S.~R., Nakar, E., et al.\ 2006, \nat, 442, 1014 
\bibitem[Sparre et al.(2011)]{2011ApJ...735L..24S} Sparre, M., Sollerman, J., Fynbo, J.~P.~U., et al.\ 2011, \apjl, 735, L24
\bibitem[Starling et al.(2011)]{2011MNRAS.411.2792S} Starling, R.~L.~C., Wiersema, K., Levan, A.~J., et al.\ 2011, \mnras, 411, 2792
\bibitem[Tagliaferri et al.(2005)]{2005Natur.436..985T} Tagliaferri, G., Goad, M., Chincarini, G., et al.\ 2005, \nat, 436, 985
\bibitem[Tanvir et al. (2012)]{Tanvir2012} Tanvir, N. R., Levan, A. J., Cucchiara, A., \& Fox, D. B. 2012, GCN Circ. 13251
\bibitem[Troja et al. (2012)]{Troja2012} Troja, E., et al., 2012, GCN Circ. 13243
\bibitem[Virgili et al. (2009)]{Virgili09} Virgili, F., Liang, E.-W., Zhang, B. 2009, MNRAS, 392, 91
\bibitem[Waxman et al.(2007)]{2007ApJ...667..351W} Waxman, E., M{\'e}sz{\'a}ros, P., \& Campana, S.\ 2007, \apj, 667, 351 
\bibitem[Wiersema et al. (2012)]{Wiersema 2012} Wiersema, K. et al., 2012, GCN Circ. 13276
\bibitem[Zhang et al.(2004)]{2004ApJ...601L.119Z} Zhang, B., Dai, X., Lloyd-Ronning, N.~M., \& M{\'e}sz{\'a}ros, P.\ 2004a, \apjl, 601, L119
\bibitem[Zhang et al. (2006)]{Zhang06} Zhang, B., et al. 2006, ApJ, 642, 354
\bibitem[Zhang et al. (2007b)]{Zhang07b} Zhang, B., Zhang, B.-B., Liang, E.-W. et al. 2007b, ApJ, 655, L25
\bibitem[Zhang et al. (2009a)]{Zhang09a} Zhang, B., Zhang, B.-B., Virgili, F. et al. 2009a, ApJ, 703, 1696
\bibitem[Zhang et al. (2007a)]{ZLZ07} Zhang, B. B., Liang, E. W., \& Zhang, B., 2007a, ApJ, 666, 1002
\bibitem[Zhang et al. (2009b)]{Zhang09b} Zhang, B. B., Zhang, B., Liang, E. W., \& Wang, X. Y. 2009b, ApJ, 690, L10
\bibitem[Zhang(2008)]{2008ApJ...685.1052Z} Zhang, F.-W.\ 2008, \apj, 685, 1052
\bibitem[Zhang et al. (2004b)]{Zhang04b} Zhang, W. Woosley, S. E., Heger, A. 2004b, ApJ, 608, 365
\end{thebibliography}
\end{document}